\documentstyle[eqsecnum,prb,aps,twocolumn,epsf]{revtex}

\begin{document}

\draft

\title{Field-induced breakdown of the quantum
Hall effect}
 \author{K. Shizuya}
  \address{Yukawa Institute for Theoretical Physics\\
 Kyoto University,~Kyoto 606-8502,~Japan }

\maketitle
 
\begin{abstract} 
A numerical analysis is made of the breakdown of the quantum Hall 
effect caused by the Hall electric field in competition with disorder.
It turns out that in the regime of dense impurities, in
particular, the number of localized states decreases exponentially
with the Hall field, with its dependence on the magnetic and electric
field summarized in a simple scaling law. The physical picture
underlying the scaling law is clarified. 
This {\it intra}\,subband process, the competition of the Hall field
with disorder, leads to critical breakdown fields of magnitude of a few
hundred V/cm, consistent with observations, and accounts for their
magnetic-field dependence $\propto B^{3/2}$ observed experimentally.
Some testable consequences of the scaling law are discussed.
\end{abstract}


\section{Introduction}

Since the discovery of the quantum Hall effect~\cite{V,Rev,AA,Pr,L}
(QHE) there has been considerable interest in 
its breakdown~\cite{EKPW,Cage,SCTH,KHS,BMSL,Kawaji,BBKL,Shimizu}
due to high currents.
Rich accumulation of experimental data has recently shed new light on
this breakdown phenomenon: A series of experiments 
by Kawaji {\it et al.}~\cite{Kawaji} 
using butterfly-shaped Hall bars clearly showed   
that the breakdown of the QHE is controlled by
the current density or the Hall field rather than the
current itself and that for samples with a variety of electron
concentration and mobility the critical values of the Hall field at
different diagonal-resistance plateaus scale like 
$\propto B^{3/2}$ with the magnetic field $B$.
This $B^{3/2}$ dependence disfavors some of the breakdown mechanisms 
proposed earlier, such as those attributed to electron
heating~\cite{EKPW} and phonon emission.~\cite{SCTH}  
Zener tunneling~\cite{SCTH,TDG,HTG,ES,Trugman} of electrons 
across the Landau gap appears consistent with it but theoretically
yields critical fields one order of magnitude larger than observed
experimentally. 

In a previous paper~\cite{KSns} we presented  
a numerical study of current distributions in Hall bars.
There we noted, as a by-product of the study, the possibility that
the Hall field in competition with disorder in the
sample interior would be responsible for the breakdown of the QHE,
with a simple estimate of the competition that leads to critical
fields of magnitude consistent with observations and 
their $B^{3/2}$ scaling law. 
The purpose of the present paper is to verify by a detailed numerical
analysis that this intrasubband process can indeed account for the
breakdown phenomenon.
Actually Stormer {\it et~al.}~\cite{SCTH} earlier noted that the action 
of a Hall field could qualitatively explain many of their experimental
findings.  Recently Shimada {\it et al.}~\cite{Shimizu} also
suggested, through a measurement of activation energy, that 
the $B^{3/2}$ scaling law is related to field-dependent broadening of
the spectra of extended states. 
We shall study numerically how the fraction of 
delocalized states per Landau subband depends on the electric and
magnetic field and clarify the physical picture underlying the
scaling law.

In Sec.~II we review a theoretical framework~\cite{KSns} for our
numerical simulations.
We examine the case of a single impurity in Sec.~III and handle Hall
samples with random impurities in Sec.~IV. 
Section~V is devoted to a summary and discussion.

\section{Formalism}

Consider electrons confined to a strip of length $L_{x}$ and width
$L_{y}$ (or formally, a strip bent into a loop of circumference
$L_{x}$), described by the Hamiltonian:
\begin{eqnarray}
&&{\sf H} =H_{0}+ U(x,y)-eA_{0}(y), \label{HAU}\\
&&H_{0} = {1\over 2}\omega \Bigl\{ \ell^{2}p_{y}^{2}
+(1/\ell^{2})(y - y_{0})^{2}\Bigr\},
 \label{hzero}
\end{eqnarray}
written in terms of $\omega \equiv eB/m$,
the magnetic length ${\ell}\equiv 1/\sqrt{eB}$ and 
$y_{0} \equiv p_{x}/(eB)$. 
Here a uniform magnetic field $B$ normal to the plane has been
introduced by use of the Landau-gauge vector potential $(-By,0)$.

We shall detect the current $j_{x}(x,y)$ flowing in the presence of a
Hall potential  $A_{0}(y)$ and an impurity potential
\begin{equation}
U(x,y) = \sum_{i}\lambda_{i}\, \delta (x - x_{i}) \delta (y - y_{i}),
 \label{Uxy}
\end{equation}
which consists of short-range impurities of strength $\lambda_{i}$
distributed over the sample at position $(x_{i},y_{i})$.
For simplicity, we here take $A_{0}(y)= - y E_{y}$ which  
produces a uniform Hall field $E_{y}$ in the $y$ direction.

The cyclotron motion of an electron is described by $H_{0}$, whose 
eigenstates are Landau levels $|N\rangle = |n, y_{0}\rangle$ labeled
by integers $n = 0,1,2,\cdots$, and $y_{0}=\ell^{2}\,p_{x}$,
with wave functions of the form
\begin{equation}
\langle x,y |N\rangle = 
(L_{x})^{-1/2}\,e^{ixp_{x}}  \phi_{n} (y;y_{0}). \label{psiN}
\end{equation}
The $\phi_{n} (y;y_{0})$ are highly localized around $y\sim y_{0}$ 
with spread $\triangle y \sim O(\ell)$, and are given by the usual
harmonic-oscillator wave functions for electrons in the sample bulk.
In what follows we shall take no explicit account of the electron 
edge states, which, being always extended, have little to do with 
field-induced delocalization.

With $\langle x,y|N\rangle$ taken to be periodic in $x$,
$y_{0}=\ell^{2}p_{x}$  is labeled by integers $k$:
\begin{equation}
y_{0}= (2\pi\ell^{2}/L_{x}) k \equiv y_{0}[k]. 
\label{yzerok}
\end{equation}
We shall henceforth use this discrete label $k$
rather than $y_{0}$ to refer directly to each electron state, with 
normalization $\langle N|N'\rangle = \delta_{nn'}\delta_{kk'}$.

For numerical analyses it is advantageous to handle the
Hamiltonian $H_{NN'}\equiv \langle N|{\sf H}|N'\rangle$ 
in $N=(n,k)$ space:
\begin{eqnarray}
H_{N N'}\!
&=&  \omega ( n\!+{\textstyle {1\over{2}}} )\,
\delta_{NN'}\!+ U_{NN'}\! +eE_{y}\, y_{NN'},
\label{HNN}   \\
U_{NN'}\!&=&\! {1\over{L_{x}}} 
\sum_{i} \lambda_{i} e^{-i(y_{0}\! - y'_{0}) x_{i}/\ell^{2}}\! 
\phi_{n}(y_{i};\!y_{0}) \phi_{n'}(y_{i};\!y'_{0}),
\label{UNN} \\
y_{NN'}&=& \{y_{0}\,\delta_{nn'} 
+ \ell\,Y_{n n'} \}\, \delta_{k k'}  , 
\label{yNN} \\
Y_{nn'}&=& \sqrt{n'/2}\, \delta_{n+1,n'} 
+ \sqrt{n/2}\, \delta_{n,n'+1},
\end{eqnarray}
where $y_{0}=y_{0}[k]$ and $y'_{0}=y_{0}[k']$.

When the disorder $U_{NN'}$ is weak compared with the level gap,
i.e., $|\lambda_{i}/2\pi\ell^{2}| < \omega$, one can diagonalize 
$H_{NN'}$ with respect to Landau-level labels $(n,n')$ 
by a suitable unitary transformation $H^{W}= WHW^{-1}$. 
In particular, a simple $O(U)$ choice $W=e^{i\Lambda}$ with
\begin{equation}
i\Lambda_{NN'}= {e E_{y}\,Y_{nn'}\delta_{k k'}+ U_{NN'}
\over{\omega (n - n') }}  \ \ \ \ 
({\rm for\ }n\not=n') ,
 \label{lamNN}
\end{equation}
leads to the Hamiltonian
$\langle N|H^{W}|N'\rangle =\delta_{nn'}H^{W}_{nn}(k,k')$ governing
each impurity-broadened Landau subband:
\begin{eqnarray}
H^{W}_{nn}(k,k')
=&&\! \Bigl\{ \omega (n\!+{\textstyle {1\over{2}}} ) + 
eE_{y}\, y_{0} \Bigr\}\, \delta_{k k'} \nonumber \\
&& + U_{nn}(k,k') + O(U^{2}/\omega),
\label{HWnn}
\end{eqnarray}
where we have set $U_{NN'}\rightarrow U_{nn'}(k,k')$.

Numerically diagonalizing $H^{W}_{nn}(k,k')$, one obtains eigenstates
$|\alpha\rangle$ forming the $n$th subband, with wave functions of the
form $\langle n,k|\alpha\rangle$. 
Then the current density carried by an eigenstate $|\alpha\rangle$ is 
calculated through
\begin{equation}
j_{x}^{(\alpha)}(x,y) = 
e\omega\, {\rm Re} \left[ \langle \alpha|W| {\bf x}\rangle 
\langle {\bf x}| (y-y_{0}) W^{-1}|\alpha\rangle \right].
 \label{jxalpha}
\end{equation}

Impurities capture orbiting electrons and make them localized in
space. In contrast, an applied (or internally generated) Hall field
causes a drift of electrons with velocity $v_{x}\sim E_{y}/B$ and
works to delocalize them. A simple estimate of energy cost~\cite{KSns}
reveals how these two effects compete:
Consider an impurity of strength $\lambda$ placed at
$x=y=0$ on an infinite plane with $E_{y}=0$.
It traps an electron and in each Landau level $n$ there arises one
localized state of energy shift 
$\triangle \epsilon \approx \lambda/(2\pi \ell^{2})$ relative to the
level center $\omega (n + {\textstyle {1\over2}})$. 
[This is readily seen if one notes that $U_{n n}(k,k')$ in the present 
case is a projection to the $n$th level 
$(\propto |n\rangle \langle n|)$ of
a harmonic oscillator with coordinate $y_{0}$.] 
It is thus convenient to define a dimensionless strength $s$
by $\lambda/(2\pi \ell^{2}) = s\, \omega$.
As a Hall field is turned on, this localized state will perceive 
over its spatial extent of $O(\ell)$ an energy variation of 
magnitude $\sim e\ell\,E_{y}$.
Accordingly, if the field becomes so strong that 
\begin{equation}
e \ell\, |E_{y}| \gtrsim |\lambda|/(2\pi \ell^{2}) =|s|\, \omega,
\label{svsE}
\end{equation}  
the impurity would fail to capture any electron. 
In the next section we shall make this criterion for
delocalization quantitative by a numerical analysis .

\section{The case of a single impurity}
Consider a sample of length 
$L_{x}=\sqrt{2\pi}\ell \times 16 \approx 40 \ell$ and width
$L_{y}=\sqrt{2\pi}\ell \times 8 \approx 20 \ell$,
supporting $16\times 8=128$ electron states per level with 
$-{1\over2} L_{y} \le y_{0} < {1\over2}L_{y}$, 
or $y_{0}=(2\pi\ell^{2}/L_{x}) k$ with $k=-64, -63,\cdots, 63$.
Place an impurity of strength $s=0.1$ at $(x,y) = (10\ell,0)$ 
in the middle of the sample and examine the electron states about the 
impurity by varying the field $E_{y}$ or the ratio
\begin{equation}
R=e \ell\, |E_{y}|/ s\omega .
 \label{Ratio}
\end{equation}
It turns out that the electrons states near the boundary 
$y = \pm {1\over2} L_{y} \sim \pm 10 \ell$ are hardly affected
by the impurity.

A close look into the $n=0,1,2,3$ levels reveals the following 
general features:
In the $n$th level there arise $(n+1)$ states of sizable spread 
(in $y_{0}$) $\sim O(\ell)$, all of which turn out
localized when $E_{y}$ is sufficiently weak (e.g., $R<0.001$).
Only one of them, having
$\triangle \epsilon \approx s \omega$, is tightly localized. 
The other $n$ states gain practically no energy shift 
$\triangle \epsilon \approx 0$ and get easily delocalized as $E_{y}$
is increased, in accordance with the criterion~(\ref{svsE}).

In what follows we shall focus on the tightly-localized states
with $\triangle \epsilon \approx s\,\omega$; 
there is one such state per level. 
They carry no current while $E_{y}$ is sufficiently weak. 
For $R > 0.15$ they start carrying an appreciable amount of current
$J_{x}^{\rm loc} = \int dy\, j_{x}^{\rm loc}(x,y)$, 
increasing with $E_{y}$ on average and becoming sizable at some
discrete values of $E_{y}$. 
Figure~1 shows such current-field characteristics for a
potentially-localized state in the $n=0$ level, 
along with a smooth curve interpolating through the local minima.

A direct look into current distributions shows that for these discrete 
values of $E_{y}$ a localized state strongly interferes with an
extended state nearly degenerate in energy and lying within its
spatial extent. Noting that extended states are equally spaced with
separation $\triangle y_{0}\equiv 2\pi\ell^{2}/L_{x}$ in $y_{0}$
space, one expects that such interference takes place
when $s\,\omega \approx e E_{y} \triangle y_{0} \times$\,integer,
i.e., at discrete values of the ratio $R$,
\begin{equation}
R^{\rm disc} \approx (\ell/\triangle y_{0}) / K
\label{Rdisc}
\end{equation}
with some integers $K=1,2,\cdots$.
Indeed, the sequence 
\begin{equation}
R^{\rm disc}\approx 0.1705, 0.1803, 0.1913, 0.204,\cdots,
 \label{Rseq}
\end{equation}
obtained for the localized state of the $n=0$ level 
is reproduced from Eq.~(\ref{Rdisc}) 
with $\triangle y_{0}\approx 0.3133 \ell$ and 
$K=19,18,17,\cdots$ to accuracy of two percent or less.
Agreement becomes almost perfect if one notes that an $O(E_{y}^{2})$
correction makes the energy $\triangle \epsilon^{\rm loc}$ of 
the localized state larger than 
$\lambda/(2\pi \ell^{2})$ by factor $f=1 + {1\over2}R^{2}$ and
includes this factor on the right-hand side of Eq.~(\ref{Rdisc}).

The smooth interpolating current-field curve in Fig.~1 rises 
from $10^{-6}$ to $10^{-2}$ as $R$ varies from 0.23 to 0.33; 
this shows that delocalization proceeds abruptly 
over a small range of $R$ or $E_{y}$. 
Actually, a simple parameterization 
$J_{x}^{\rm loc}= -e (E_{y}/B)(1/L_{x}) F_{n=0}(R)$ 
with 
\begin{equation}
	F_{n=0}(R) = 1/\exp[\exp[5.264  - 11.52 R]]
	\label{fitFn}
\end{equation}
gives a good fit to the curve for $0.17\lesssim R \lesssim 0.33$.

Because the onset of the interference is determined geometrically,
the sequences $R^{\rm disc}$ for analogous localized states 
of higher levels are essentially the same as the $n=0$ sequence,
with slight deviations (of a few percent) in overall magnitude
caused by the $O(E_{y}^{2})$ correction to 
$\triangle \epsilon^{\rm loc}$, which for the $n$th level is given by
factor $f_{n}=1+ (n+ {1\over2})R^{2}$.  
The interference, however, becomes stronger with $n$ (because
of larger spatial overlap among nearly-degenerate states) and  
the interpolating current-field curves rise 
more steeply with $R$, as shown in Fig.~1 for the $n=1,2,3$ levels.
These curves are distinct, but almost coincide by simple rescaling 
of the argument:
\begin{eqnarray}
F_{n=0}(R) &\approx& F_{n=1}(R/1.07) \nonumber\\
&\approx& F_{n=2}(R/1.11) \approx F_{n=3}(R/1.15) ,
	\label{Fn}
\end{eqnarray}
which holds quite well for $0.001 \lesssim F_{n} \lesssim 0.01$.
One may define a critical ratio $R^{\rm cr}$ as such 
at which $F_{n}(R)$ reaches some given value, say, $1/10^{3}$.
Then, Eq.~(\ref{Fn}) implies that $R^{\rm cr}$ decreases only 
by a few percent as one moves to the adjacent higher level;
indeed, from Fig.~1 one can read off 
$R^{\rm cr} \approx (0.29,0.27,0.26,0.25)$ for $n=(0,1,2,3)$.
This shows that 
$R= e\ell |E_{y}|/(s\omega)$ is a good measure to express 
field-induced delocalization of electron states, 
with only slight reference to specific levels $n$.

We have made the sample twice and four times longer
$(L_{x} \rightarrow 2 L_{x} \rightarrow 4 L_{x})$ and 
observed that $F_{n}(R)$ scale with $L_{x}$ in such a 
way that $(1/L_{x}) F_{n}(R)$ stay fixed for each $n=0,1,2,3$
over the range $0.17\lesssim R \lesssim 0.32$, giving rise to 
a relation analogous to Eq.~(\ref{Fn}). It thus appears that
$F_{n=0}(R)$ has a universal meaning of its own.

\section{Case of random impurities}

Let us next consider samples with random impurities. 
We earlier considered~\cite{KSns} 
180 short-range impurities of varying strength 
within the range $ |s_{i}| \le 0.1$, randomly distributed over a
domain of length $\sqrt{2\pi}\ell \times 28 \approx 70 \ell$ 
and width $\sqrt{2\pi}\ell \times 6 \approx 15 \ell$. 
Here we shall reshuffle them and generate a new configuration of 
360 impurities distributed over a domain of larger width 
$\approx 30 \ell$; see the appendix for details. 
We embed this impurity configuration in the region 
$-15\,\ell \le y \le 15\,\ell$ of a Hall sample of length 
$L_{x}=\sqrt{2\pi}\ell \times 28 \approx 70 \ell$ and width
$L_{y}=\sqrt{2\pi}\ell \times 18 \approx 45 \ell$, 
supporting $N_{s}= 28\times18=504$ electron states per subband with 
$-{\scriptstyle 1\over2}L_{y}\le y_{0} < {\scriptstyle 1\over2}L_{y}$.
Note that the disordered domain accommodates roughly 
$N_{s}^{\rm eff}\approx 28\times 12 = 336$ electrons per subband.

Of our main concern are the electron states residing over the
disordered domain, that are to simulate electrons in the bulk of 
a realistic sample. 
We take no explicit account of the edge states and simply leave
the two zones $15\,\ell < |y| \lesssim 22.5\,\ell$ of the sample
impurity-free. 
In practice, to determine the electron states in the disordered bulk 
one has to numerically diagonalize the Hamiltonian for a sample of 
somewhat larger width like this.

For each subband $n$ we arrange the electron states in the order of
descending energy and use the number of vacant states, 
$N_{\rm v}$, to specify the filling of the subband.
For convenience we start with the filled subband ($N_{\rm v}= 0$) 
and study the total current this subband supports,
\begin{equation}
J_{x} = \int dy\, \sum_{\alpha} j_{x}^{(\alpha)}(x,y),
\end{equation} 
by removing electrons one by one.
In Fig.~2(a) we plot $J_{x}$ as a function of $N_{\rm v}$ 
for the $n=0$ subband and 
$R=10^{-4}, 0.001, 0.01, 0.025, 0.05, 0.1, 0.2$, where
\begin{equation}
R \equiv -e \ell\, E_{y}/(0.1\,\omega) 
\end{equation}
is now normalized using 
$|s|_{\rm max}\equiv {\rm Max}\{|s_{i}|\}=0.1$ .
There, simply for ease of exposing numerical results 
we have focused on 437 electron states whose center positions
$\langle \alpha |y|\alpha\rangle$ lie across the region
$|y|<19.6\,\ell$; the electron states residing outside are close to
undisturbed modes.

Let us first look at the $R = 10^{-4}$ case, 
where the upper and lower plateaus are clearly seen 
for $0 \le N_{\rm v}\lesssim 170$ and $270 \le N_{\rm v}$,
respectively, supporting about 330 localized states in total. 
There the linear decrease of $J_{x}$ 
for $170\lesssim N_{\rm v}\lesssim 210$ 
and $230\lesssim N_{\rm v}\lesssim 270$ is due to gradual
evacuation of extended states from the impurity-free zones 
$|y| \gtrsim 15\,\ell$.  A steep fall of $J_{x}$ around 
$N_{\rm v}\sim 220$ indicates that a large portion of the Hall current 
is carried by a small number of states residing in the sample bulk.

The plateaus shrink rapidly with increasing $|E_{y}|$, 
and become hardly visible for $R = 0.2$ in Fig.~2(a). 
Somewhat unexpectedly, the way the upper and lower plateaus
diminish with increasing $R$ is virtually the same for 
higher $n=1,2,3$ subbands as well, as shown in Fig.~2(b)
for the $n=2$ subband. 
Numerically, the number of states forming the plateaus, $N^{\rm pl}$,
decreases exponentially with $R$,
with little dependence on $n=0,1,2,3$, as seen from Fig.~2(c); 
$N^{\rm pl}\approx 330\ e^{- R/0.041}$ gives a good numerical 
fit.~\cite{fnNpl}

As seen from Fig.~2(a), a filled subband carries the same amount of
Hall current as in the impurity-free case. 
One may thus simply look into the uppermost occupied subband in
studying the dependence of the Hall current on the Hall field.

In Fig.~2(b) the plateaus are somewhat blurred when
$E_{y}$ is very weak, $R \lesssim 10^{-3}$. 
This blur or instability is due to the interference 
among nearly degenerate states and grows
prominent in higher subbands, especially in the regime of small $R$, 
where  $J_{x}$ itself gets small.
Such blur makes $N^{\rm pl}$ underestimate the actual plateau width, 
and this explains why apparent deviations from the universal 
$N^{\rm pl}-R$ characteristic develop for small $R$ in Fig.~2(c).

We have so far supposed changing the filling fraction 
$\nu=N_{e}/N_{s}$ of Landau subbands by varying electron
population $N_{e}$ with $B$ kept fixed. 
Let us next consider what would happen if one reaches higher subbands 
by reducing $B$ while keeping $N_{e}$ fixed, as commonly done in
experiments. Unfortunately it is impractical to change the length unit
$\ell=1/\sqrt{eB}$ continuously in numerical simulations.  
We shall therefore take an indirect route and examine what happens if
the magnetic field is changed from $B_{0}$ to $B_{0}/\kappa$.  
We suppose that the sample size~\cite{fnsize} is defined in units of 
$\ell_{0}=1/\sqrt{eB_{0}}$ of the original field $B_{0}$.
Then, as $B$ is varied, the disordered domain accommodates about
$N_{s}^{\rm eff}\approx 336/\kappa$ states per subband so that 
$N_{\rm imp}/N_{s}^{\rm eff}\approx \kappa$.  
Thus, reducing the magnetic field $B_{0} \rightarrow B_{0}/\kappa$ not
only attains a larger filling factor $\nu \rightarrow \kappa\, \nu$
but also effectively makes impurities $\kappa$ times denser.

Some remarks are in order here. 
We have so far counted the number of localized states through 
the plateau width $N^{\rm pl}$, which best visualizes 
how the Hall plateaus shrink with $R$.
This $N^{\rm pl}$, however, tends to overlook (a small portion of) 
localized states hidden off the plateaus and becomes less reliable 
for large $R$, where $N^{\rm pl}$ gets small. 
Fortunately there is another measure, less direct but better
suited for detecting the disappearance of Hall plateaus:
One can directly count the number $N^{\rm loc}$ of states that carry
an almost vanishing amount of current per state, e.g., less than 1\%
of the unit $(-e/L_{x})E_{y}/B$. (We adopt this choice of 1\% below.)

Figure~3(a) shows $N^{\rm loc}$ of the uppermost subband, plotted as a
function of $R \propto e\ell\,|E_{y}|/\omega$, for various choices of
$B= B_{0}/\kappa$ with $\kappa=0.33 \sim 4$.  
Actually the data refer specifically to the $n=0$ subband but, 
in view of what we have learned from the fixed-$B$ cases,
we suppose that they apply to higher subbands as well.  
One may thus regard, e.g., the $\kappa=2$ curve as 
referring to the $n=1$ subband (or the spin-split $n=0$ subband
if the electron spin is taken into account) with $1< \nu \le 2$, 
which is reached from the $\kappa=1$ curve referring to 
the $n=0$ subband with $0< \nu \le 1$ 
by reducing $B=B_{0} \rightarrow {1\over2}B_{0}$.

 Figure~3(b) tells us that, in the regime of dense impurities 
$N_{\rm imp}/N_{s}^{\rm eff} \approx \kappa \gtrsim 1$, 
the  $N^{\rm loc} - R$ curves appear quite universal, exhibiting, like
$N^{\rm pl}$, an exponential fall-off over a wide range $R \gtrsim
0.1$, with slight dependence on $\kappa$ for its range of change 
$1\le \kappa \le 4$. The apparent deviations, growing with $\kappa$ 
for $R<0.1$, are due to the interference among nearly degenerate
states. Thus, an approximate scaling law of the form 
\begin{eqnarray}
N^{\rm loc} &\approx& N_{0}\, e^{- R/0.058} ,
	\label{NlocE}  
\end{eqnarray}
with $N_{0}\approx N_{\rm imp}=360$, determines how $N^{\rm loc}$
depends on $E_{y}$ and $B$ over a wide range $R \gtrsim 0.1$.

As for the normalization $N_{0}$ it is clear that
$N_{0}\propto N_{\rm imp}$ in general, and that
$N_{0}\approx N_{\rm imp}$ when $N_{\rm imp}\approx N_{s}^{\rm eff}$,
in which case all impurities are expected to capture electrons in the
$E_{y} \rightarrow 0$ limit. 
It is remarkable that $N_{0}\approx N_{\rm imp}$ holds well even
though the ratio $N_{\rm imp}/N_{s}^{\rm eff}$ far exceeds unity
in Fig.~3(b).

The situation somewhat changes in the regime of dilute impurities,
$N_{\rm imp}/N_{s}^{\rm eff} < 1$. See Fig.~3(a) and (c).
There, as impurities become dilute ($\kappa=1 \rightarrow 0.33$),
$N^{\rm loc}$ rises more rapidly with reducing $R$ and 
tends to saturate for $R\sim 0$;
this is because all the impurities would capture
electrons before $R$ reaches zero.  
Thus, for dilute impurities $N_{\rm imp}/N_{s}^{\rm eff}\ll 1$, 
the $N^{\rm loc} - R$ characteristics significantly deviate 
from the scaling form~(\ref{NlocE}).
Still the values of $N^{\rm loc}$ in the near-breakdown regime 
($R =0.25 \sim 0.4$) stay roughly the same, and actually are
consistent with those of the $N_{\rm imp}/N_{s}^{\rm eff} > 1$ case.

A physical picture suggested by this stability of $N^{\rm loc}$
for large $R$ is that the electron states remaining localized 
in the near-breakdown regime are always governed by the same set of
impurities of large strength $|s_{i}| \sim |s|_{\rm max}$;
that is, one always encounters the same set of localized states 
in the near-breakdown regime.
If this is the case, weaker impurities ($|s_{i}| \ll |s|_{\rm max}$)
would be irrelevant to the onset of the breakdown.

To test this picture we have examined the case of a truncated
impurity configuration where only 180 impurities of strength 
$0.05 \le |s_{i}|\le 0.1$ are retained, with the weaker ones
suppressed. The resulting $N^{\rm loc} - R$ characteristics, 
plotted with unfilled symbols in Fig.~3(d) for $\kappa=1,2,3,4$,
are indeed consistent with those of the original 
$N_{0}\approx N_{\rm imp}=360$ case for $R \gtrsim 0.25$.
Further truncation of impurities into an even smaller range 
$0.075 \le |s_{i}|\le 0.1$ entails no essential change.
This supports our picture and
implies that for a general impurity configuration 
$N_{0}$ is not simply given by $N_{\rm imp}$ but by the effective
number of impurities extrapolated from the density of strong
impurities ($|s_{i}|\sim |s|_{\rm max}$).
We have also examined a number of cases with $N_{\rm imp}=180$ 
and 720, and arrived at essentially the same conclusion.

A practical way to define the disappearance of a Hall plateau is to
refer to a critical value of $E_{y}$ at which $N^{\rm loc}$ or the
ratio $N^{\rm loc}/N_{e}$ reaches a certain prescribed number.
Then, it follows from the universal $N^{\rm loc}-R$ characteristic in
the near-breakdown regime that the critical field scales like  
$e\ell\, |E_{y}^{\rm cr}|/(0.1\,\omega) =C=$ const., i.e., 
\begin{equation}
e |E_{y}^{\rm cr}| =  |s|_{\rm max}\, C\, \omega/\ell \propto B^{3/2},
\label{EvsB}
\end{equation}
where we have isolated the dependence of $E_{y}^{\rm cr}$ on the
sample-specific impurity strength $|s|_{\rm max}$.
It is understood that $|s|_{\rm max}$ now refers to a typical strength
of (strong) impurities relevant in the near-breakdown regime.~\cite{fnsi} 
The scaling law $|E_{y}^{\rm cr}| \propto B^{3/2}$ applies to each
spin-split subband reached by varying $B$.

To get an idea of the magnitude $|E_{y}^{\rm cr}|$ 
let us look at Fig.~3(b) and choose $C \approx 0.25$, 
for which we count only 10 localized states in the uppermost subband. 
This leads to a critical field of magnitude 
$|E_{y}^{\rm cr}| \sim 250$ V/cm 
at the $\nu=1$ plateau for typical values $\omega \approx 10$ meV 
and $\ell \approx 100$ \AA, with $|s|_{\rm max}=0.1$.
Critical fields of this order of magnitude 
appear to be consistent with experiments.~\cite{Kawaji}

There are some testable consequences of the scaling law.
First it is of interest to study how the Hall-plateau width depends on 
$\nu$ and $E_{y}$. Let $B^{(\nu=1)}$ denote the value of $B$ at which
the lowest subband is filled up, so that one can write the filling
factor as $\nu = B^{(\nu=1)}/B$.
Let us suppose that each subband contains the same number of 
localized states $\sim {1\over2}N^{\rm loc}$ in its upper and lower
halves of the spectrum.  
Then a plateau develops near integer filling $\nu \sim n$ 
if $|n N_{\rm s}- N_{e}| \le {1\over2} N^{\rm loc}$ is satisfied; 
i.e., the plateau lies over the range $\nu_{-}\le \nu \le \nu_{+}$ 
with $\nu_{\pm}$ determined from 
$n/\nu_{\pm}=1\mp {1\over2} N^{\rm loc}(\nu_{\pm};E_{y})/N_{e}$, 
where $N^{\rm loc}$ depends on $\nu$ and $E_{y}$.
When $E_{y}$ is very weak and localization is almost complete, i.e., 
$N^{\rm loc}\rightarrow N_{\rm s}$, this leads to
$\nu_{\pm}\approx n \pm {1\over2}$ so that the plateau width
\begin{equation}
\triangle \nu \equiv \nu_{+}-\nu_{-} \approx 1
\end{equation}
at each plateau, as observed experimentally~\cite{PTG} at low
temperatures. 
[The observation~\cite{PTG} of sharp steps connecting the plateaus
would suggest that the case of dense impurities  
$N_{\rm imp}/N_{s}>1$ applies to realistic samples.] 

In contrast, in the near-breakdown regime $N^{\rm loc}/N_{e}\ll 1$ 
(or for large $E_{y}$) 
we find from Eq.~(\ref{NlocE}) the plateau width
$\triangle B/B^{(\nu=1)}\equiv 1/\nu_{+} - 1/\nu_{-}$ of the form
\begin{equation}
\triangle B \approx {1\over{n}}\,B^{(\nu=1)}(N_{0}/N_{e})\, 
e^{ - \xi\, n^{3/2}E_{y}}
 \label{triB}
\end{equation}
around $\nu \sim n$, where 
$\xi=(C/0.058)/E_{y}^{\rm cr\ (\nu \sim 1)}$ refers to the critical
field $E_{y}^{\rm cr\ (\nu \sim 1)}$ at the $\nu=1$ plateau.
Thus, near the breakdown the width $\triangle B$ scales like
$(1/n)e^{ - \xi\, n^{3/2}E_{y}}$ at $\nu \sim n$ plateaus. 
A close look into how the width $\triangle B$ decreases with $E_{y}$ 
at different plateaus will reveal another manifestation of 
the $E_{y}^{\rm cr} \propto B^{3/2}$ law.

The number of visible plateaus is readily determined also.
Suppose that the total current $J_{x}$ is kept fixed while $B$ is
varied, as normally done in experiments.
Let $J_{x}^{\rm cr\ (\nu \sim 1)}$ denote the critical current at 
which the $\nu \sim 1$ plateau disappears. Then the critical current
at the $\nu \sim n$ plateau scales with $n$ like 
$J_{x}^{\rm cr\ (\nu \sim n)} = n^{-1/2} J_{y}^{\rm cr\ (\nu \sim1)}$.
As a result, for given $J_{x}$ one will observe $n_{\rm max}$
plateaus, while reducing $B$, where $n_{\rm max}$ denotes the largest
integer satisfying the equation 
\begin{equation}
 n<  (J_{x}^{\rm cr\ (\nu \sim 1)}/ J_{x})^{2}.
\end{equation}

\section{Summary and discussion}
 
In this paper we have studied the competition of the Hall field with
disorder in triggering the breakdown of the QHE.  
In the case of a single impurity we have seen that field-induced
delocalization of a localized state starts with
its interference with nearly-degenerate extended states.
This geometrical nature of the onset and subsequent abrupt 
growth make the delocalization process, when expressed in terms of 
$e\ell E_{y}/B$, practically independent of the Landau-subband index
$n$.  Such $n$ independence is largely carried over to
the case of random impurities. 
We have seen, in particular, that in the regime of dense impurities 
$N_{\rm imp}/N_{s}>1$ (which presumably applies to realistic samples)
the number of localized states, decreasing exponentially with
$|E_{y}|$, obeys an approximate scaling law written in terms of 
$e\ell E_{y}/\omega \propto  E_{y}/B^{3/2}$.
Deviations from the scaling form arise in the regime of dilute 
impurities. Nevertheless the results of our numerical analysis are
neatly summarized in a physical picture that 
the breakdown is controlled by a fixed set of impurities of large
strength $|s_{i}| \sim |s|_{\rm max}$, that support localized states
in the near-breakdown regime. 
This picture offers a simple explanation for 
the empirical $E^{\rm cr}_{y} \propto B^{3/2}$ law of 
the critical breakdown field, and leads to some testable consequences
concerning the number and widths of visible Hall plateaus.
It is also consistent with the observation of Cage 
{\it et al.}~\cite{Cage} that the breakdown of the dissipationless 
current is spatially inhomogeneous.

Experimentally the $E^{\rm cr}_{y} \propto B^{3/2}$ law is 
established not only for each specific sample but also for samples 
differing in carrier density and mobility.~\cite{Kawaji}
This experimental fact indicates that the critical field
$E_{y}^{\rm cr}$ is practically independent of both 
electron concentration $\propto N_{e}$ and 
impurity concentration $\propto N_{\rm imp}$.  
This is consistent with the abovementioned physical picture or its
outcome in Eq.~(\ref{EvsB}), where the critical field, though
sensitive to the characteristic strength $|s|_{\rm max}$ of disorder,
appears practically independent of $N_{e}$ and $N_{\rm imp}$.
The $|s|_{\rm max}$ is related to the width of each subband.
Its magnitude $|s|_{\rm max} \sim O(0.1)$ differs
from a sample to another, but its range of variation is presumably 
of $O(1)$, far smaller than the ranges of variations
of sample-specific electron and impurity concentrations.
The observed $E^{\rm cr}_{y} \propto B^{3/2}$ law would thus be
attributed to relatively small variations in $|s|_{\rm max}$ for
a variety of samples. 
It is also a natural consequence of the present {\it intra}\,subband
process that the magnitude of $E^{\rm cr}_{y}$ falls within the
observed range of a few hundred V/cm, one order of magnitude smaller
than what intersubband processes~\cite{TDG,HTG,ES} typically predict.

We have used a uniform field $E_{y}$ to detect the current-carrying
characteristics of Hall electrons.  In real samples the Hall field 
generated by an injected current is not necessarily uniform and has
actually been observed to be stronger 
near the sample edges.~\cite{FKHBW} 
Under such general circumstances the field $E_{y}$ employed in
our analysis should refer to the smallest of local averages of the
Hall field in the sample bulk.

\acknowledgments

The author wishes to thank S. Kawaji for useful discussions.
This work is supported in part by a Grant-in-Aid for 
Scientific Research from the Ministry of Education of Japan, 
Science and Culture (No. 10640265).  \\

\appendix
\section{Impurity distribution}
In this appendix we describe how to construct the configuration of 360
random impurities used in Sec.~IV.
We earlier used a configuration 
P$_{180}=\{ (\bar{x}_{i},\bar{y}_{i},s_{i}), i=1,\cdots,180\}$
of 180 impurities of strength $s_{i}$ located at
$(\bar{x}_{i},\bar{y}_{i})$
randomly distributed over the range $0\!\le\!\bar{x}\!\le\!24$, 
$0\!\le\!\bar{y}\!\le\!6$ and $|s_{i}| \le 0.1$.
Generate out of this P$_{180}$ another set 
P'$_{180}=\{ (\bar{x}_{i+180},\bar{y}_{i+180},s_{i+180}), 
i=1,\cdots,180\}$  via a shift (10,2,0.08) [which is our arbitrary
choice] and rearrangement so that
$\bar{x}_{i+180}=\bar{x}_{i} +10$ mod 24, 
$\bar{y}_{i+180}=\bar{y}_{181-i}+ 2$ mod 6, and
$s_{i+180}=s_{i}+ 0.08$ mod 0.2, 
within the range $0\!\le\!\bar{x}\!<\!24$, 
$0\!\le\!\bar{y}\!<\!6$ and $|s_{i}| \le 0.1$.

Then shift P$_{180}$ by (0, - 6, 0) and combine it with P'$_{180}$ to
form a set P$_{360}$ of 360 impurities distributed over the wider
range $0\!\le\! \bar{x}\!\le\! 24$ and $-6\!\le\!\bar{y}\!\le\! 6$. 
Finally, via rescaling $x_{i}=\sqrt{2\pi}\ell \times (28/24)
\times \bar{x}_{i}$ and $y_{i}=\sqrt{2\pi}\ell \times \bar{y}_{i}$ 
we distribute these 360 impurities at positions $(x_{i},y_{i})$ over
the domain $0\le x \lesssim 70\,\ell$ and 
$-15\,\ell \lesssim y \lesssim 15\,\ell$.


\newpage


\begin{figure}
\epsfxsize=8.cm
\centerline{\epsfbox{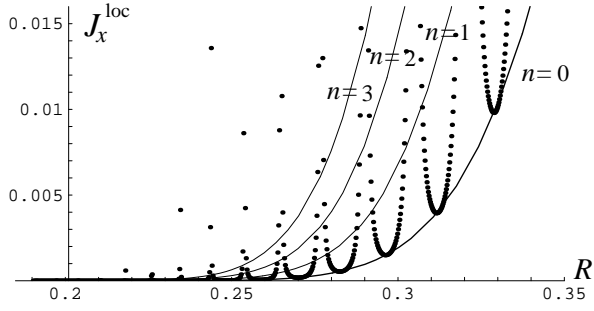}}
\vspace{0.15in}
\caption{A potentially localized state of the $n=0$ level carries an
amount of current $J_{x}^{\rm loc}$ [plotted in units of 
$-(e/L_{x}) E_{y}/B$], that becomes sizable at some discrete values of
$R= e\ell |E_{y}/s\omega|$, with a smooth curve interpolating
through the local minima.  
Also shown are analogous interpolating current-field curves for the
localized states of the $n=1,2,3$ levels.} 
\end{figure}


\begin{figure}
\epsfxsize=8.5cm
\centerline{\epsfbox{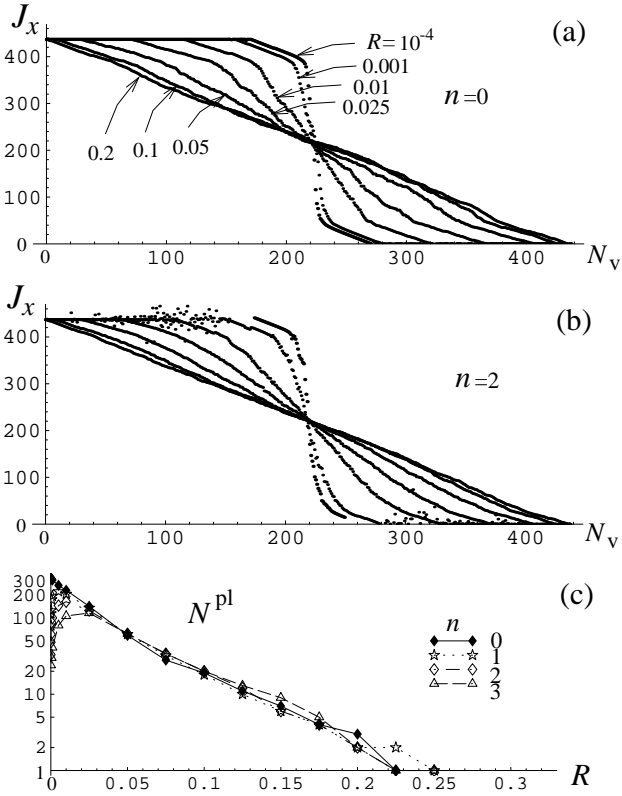}}
\vspace{0.15in}
\caption{ (a) The Hall current $J_{x}$ carried by the $n=0$ subband
[in units of $-(e/L_{x})E_{y}/B$] vs vacancy $N_{\rm v}$ for 
$R\equiv e\ell\, |E_{y}|/|s|_{\rm max}\omega = 0.0001, 0.001, 0.01,
0.025, 0.05, 0.1, 0.2\,.$\\
(b) $J_{x}$ vs $N_{\rm v}$ for the $n=2$ subband.  
The blur about the plateaus shown is associated 
with the $R=0.001$ case, and the $R=10^{-4}$ curve is only partially
shown.\\
(c) The plateau width $N^{\rm pl}$ decreases exponentially with $R$,
as shown in a $\log N^{\rm pl}$ vs $R$ plot for
the $n=0,1,2,3$ subbands.  
The lines joining the data points are meant to guide the eye.}
\end{figure}

\begin{figure}
\epsfxsize=8.5cm
\centerline{\epsfbox{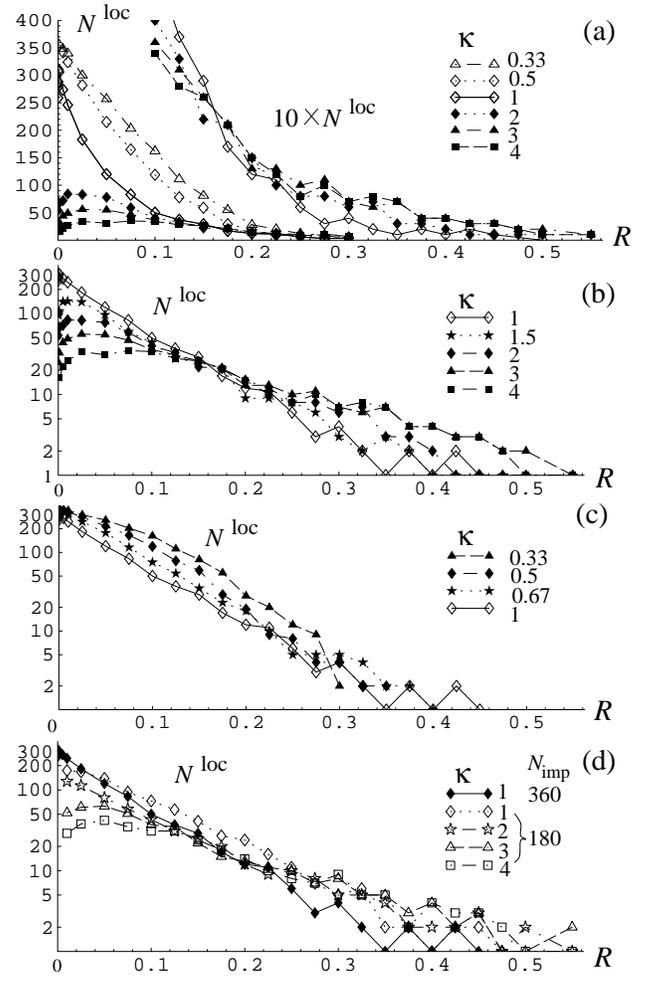}}
\vspace{0.15in}
\caption{The number of localized states per subband, $N^{\rm loc}$, 
plotted as a function of $R=e\ell\,|E_{y}|/|s|_{\rm max}\, \omega$.\\ 
(a) $N^{\rm loc}$ vs $R$ with $\kappa = 0.33 \sim 4$;
$10\times N^{\rm loc}$ shown for $R\ge 0.1$. \\
(b) $\log N^{\rm loc}$ vs $R$ with $\kappa= 1, 1.5, 2, 3, 4$.\\
(c) $\kappa=1, 0.67, 0.5, 0.33$. \\
(d) $\log N^{\rm loc}$ vs $R$ for a truncated configuration with only
180 impurities of srtength $0.05\le |s_{i}|\le 0.1$ kept.} 
\end{figure}


\begin{thebibliography}{99}
\bibitem{fnNpl}
To define $N^{\rm pl}$ we have allowed a small deviation and counted
the number of states that satisfy the cutoff 
$|J_{x}-{\rm (plateau\ values)}|< 0.5\,$unit of $(-e/L_{x})E_{y}/B$.
\bibitem{fnsize} 
In units of $\sqrt{2\pi}\ell$ the sample size is
$(28/\sqrt{\kappa}) \times (18/\sqrt{\kappa})$, which we approximate
by some appropriate integers so that the number of states $N_{\rm s}$
takes an integer. In what follows, 
we actually take $28/\sqrt{\kappa}\approx (23,20,16,14)$
and $18/\sqrt{\kappa}\approx (15,13,12,10)$ for 
$\kappa$ =(1.5, 2, 3, 4), respectively.
Similarly, for $\kappa$ =(0.67, 0.5, 0.33) we take the length to be
(34,40,48) and width to be (20,22,26), respectively.
\bibitem{fnsi}
Note here that 
$s_{i}=\lambda_{i}/(2\pi\ell^{2}\omega) = \lambda_{i}\,m^{*}/(2\pi)$ 
are independent of the magnetic field $B$ 
(as long as $\lambda_{i}$ are).
\end{thebibliography}
\end{document}